\documentclass[amsmath,amssymb,twocolumn]{revtex4}
\usepackage{graphicx}
\usepackage{amscd,amsmath,amsthm,amsfonts,amssymb}
\def\PT{$\cal{PT}$}
\def\[{\begin{equation}}
\def\]{\end{equation}}
\advance\textheight -16mm

\begin{document}
\title{Stability of soliton families in nonlinear Schr\"odinger equations with non-parity-time-symmetric complex potentials}
\author{Jianke Yang and Sean Nixon}
\address{Department of Mathematics and Statistics, University of Vermont, Burlington, VT 05401, USA}

\begin{abstract}
Stability of soliton families in one-dimensional nonlinear
Schr\"odinger equations with non-parity-time (\PT)-symmetric complex
potentials is investigated numerically. It is shown that these
solitons can be linearly stable in a wide range of parameter values
both below and above phase transition. In addition, a
pseudo-Hamiltonian-Hopf bifurcation is revealed, where pairs of
purely-imaginary eigenvalues in the linear-stability spectra of
solitons collide and bifurcate off the imaginary axis, creating
oscillatory instability, which resembles Hamiltonian-Hopf
bifurcations of solitons in Hamiltonian systems even though the
present system is dissipative and non-Hamiltonian. The most
important numerical finding is that, eigenvalues of linear-stability
operators of these solitons appear in quartets $(\lambda, -\lambda,
\lambda^*, -\lambda^*)$, similar to conservative systems and
\PT-symmetric systems. This quartet eigenvalue symmetry is very
surprising for non-\PT-symmetric systems, and it has far-reaching
consequences on the stability behaviors of solitons.
\end{abstract}


\maketitle

\section{Introduction}
Parity-time (\PT) symmetry is currently at the forefront of research
in physics and applied mathematics (see \cite{Kivshar_review,
Yang_review} for reviews). This concept started out in quantum
mechanics, where it was observed that complex potentials with
parity-time symmetry could still exhibit all-real spectra even
though the underlying Schr\"odinger operator is non-Hermitian
\cite{Bender1998}. Later, this concept spread to optics, where it
was realized that optical waveguides with even refractive-index
profiles and odd gain-loss distributions constitute \PT-symmetric
systems \cite{Musslimani2008}. In this optical setting, \PT symmetry
was observed for the first time
\cite{Segev2010,PT_lattice_exp,Peng2014}. In addition, it has been
introduced into many other physical disciplines such as
Bose-Einstein condensates, electronic circuits and mechanical
systems \cite{Longhi_2010,coupler1,Kottos2011,
BECPT2008,BECPT2012,Bender2013}. \PT systems feature a unique
property --- phase transition, where the linear spectrum changes
from all-real to partially-complex when the system parameters cross
a certain threshold
\cite{Bender1998,Ahmed2001,Musslimani2008,Nixon2012}. This phase
transition has led to interesting applications such as single-mode
\PT lasers and unidirectional reflectionless optical devices
\cite{Feng2013,Zhang2014,Mercedeh2014}. A surprising property of \PT
systems is that, even though they are dissipative due to the gain
and loss, they exhibit many properties of conservative systems, such
as all-real linear spectra and continuous families of stationary
nonlinear modes
\cite{Bender1998,Ahmed2001,Musslimani2008,coupler1,Wang2011,
Konotop2011,Zezyulin2012a, Nixon2012, Christodoulides_Bragg,
Panos_Dmitry2013,Wimmer2015,Kartashov2014}. Thus, \PT systems break
the boundary between conservative and dissipative systems and offer
novel wave-guiding possibilities. In addition, \PT systems make loss
useful, which is enlightening.

The downside of \PT symmetry stems from the restrictive conditions
set on the gain-loss profile, which must be odd. To overcome this
restriction, non-\PT-symmetric dissipative systems sharing the
properties of \PT-symmetric systems have been pursued. For instance,
wide classes of non-\PT-symmetric potentials with all-real spectra
were reported in \cite{Cannata1998, Andrianov99, SUSY2013,
Tsoy2014,PRA2016}. In addition, it was discovered that in a certain
class of such potentials with the form $g^2(x)+ig'(x)$, where $g(x)$
is an arbitrary real function, solitons also appear as continuous
families, which is very counter-intuitive
\cite{YangPLA14,Tsoy2014,Konotop_OL2014,SAPPM2015}. Furthermore, it
was argued in \cite{SAPPM2015} that potentials of the form
$g^2(x)+ig'(x)$ are the only one-dimensional (1D) non-\PT-symmetric
potentials which support soliton families. However, stability
properties of these soliton families are still largely unknown,
except for some evolution simulations of perturbed simple-shaped
solitons in a certain non-\PT-symmetric potential below a phase
transition in \cite{Tsoy2014}, which suggest that those simple
solitons could be stable.

In this paper, we systematically study the linear stability of
various soliton families in 1D nonlinear Schr\"odinger (NLS)
equations with non-\PT-symmetric complex potentials both below and
above phase transition. This study is performed by numerically
computing the linear-stability spectra of these solitons. We show
that both simple-shaped and multi-humped soliton families can be
linearly stable in a wide range of parameter values below and above
a phase transition. In addition, a pseudo-Hamiltonian-Hopf
bifurcation is revealed, where pairs of purely-imaginary eigenvalues
in the linear-stability spectra of solitons collide and bifurcate
off the imaginary axis, creating oscillatory instability, which
resembles Hamiltonian-Hopf bifurcations of solitons in Hamiltonian
systems even though the present system is non-Hamiltonian. Our most
important numerical finding is that, eigenvalues of the
linear-stability operator of these solitons appear in quartets
$(\lambda, -\lambda, \lambda^*, -\lambda^*)$, similar to
conservative systems and \PT-symmetric systems. This quartet
eigenvalue symmetry is very surprising for non-\PT-symmetric
dissipative systems, and its consequences on the linear stability of
these solitons are explained.

\section{Preliminaries} \label{s:sec2}
The mathematical model we consider is the following potential NLS
equation
\begin{equation}
i \Psi_t + \Psi_{xx} +V(x) \Psi + \sigma |\Psi|^2 \Psi = 0,
\label{e:NLS}
\end{equation}
where $V(x)$ is a complex potential, and $\sigma=\pm 1$ is the sign
of nonlinearity. This model governs nonlinear light propagation in
an optical medium with gain and loss
\cite{Kivshar_book,Yang_book,Musslimani2008}, as well as dynamics of
Bose-Einstein condensates in a double-well potential with atoms
injected into one well and removed from the other well
\cite{BEC_book,BECPT2008,BECPT2012}. If the potential $V(x)$ is
real, Eq. (\ref{e:NLS}) is conservative and Hamiltonian, and its
properties have been investigated in numerous articles for many
decades \cite{Kivshar_book,Yang_book}. If $V(x)$ is complex but
\PT-symmetric, i.e., $V^*(x)=V(-x)$, where the superscript *
represents complex conjugation, then this \PT-symmetric system has
been heavily studied in the last eight years
\cite{Kivshar_review,Yang_review}. If $V(x)$ is complex and
non-\PT-symmetric, this equation is currently at the frontier of
research. For non-\PT-symmetric potentials of the form
\[  \label{e:V}
V(x)=g^2(x)+2\gamma g(x)+ig'(x),
\]
where $g(x)$ is a real asymmetric function and $\gamma$ a real
constant, the linear spectrum of the potential can be all-real,
which is unusual \cite{Tsoy2014,PRA2016}. Note that this form of the
potential is equivalent to $g^2(x)+ig'(x)$ under a shift
$g(x)+\gamma \to g(x)$ and a gauge transformation to
Eq.~(\ref{e:NLS}). It is used in this article since it is more
convenient to induce a phase transition by varying the parameter
$\gamma$ while keeping the function $g(x)$ fixed. A more important
phenomenon with the potential (\ref{e:V}) is that, Eq. (\ref{e:NLS})
under this potential admits continuous families of solitons
\cite{Tsoy2014,Konotop_OL2014,SAPPM2015}. This is surprising since,
in typical dissipative systems, solitons exist as isolated solutions
with discrete power levels due to the requirement of balance between
gain and loss \cite{Akhmediev_book}. Dissipative but \PT-symmetric
systems admit soliton families with continuous power levels, which
is interesting \cite{coupler1,Wang2011,Konotop2011, Zezyulin2012a,
Nixon2012, Christodoulides_Bragg, Panos_Dmitry2013,Kartashov2014}.
However, the existence of such soliton families can be easily
understood due to the \PT symmetry, which assures the balancing of
gain and loss for all \PT-symmetric solitons \cite{Yang_SAPM2014}.
Soliton families in non-\PT-symmetric systems, on the other hand,
are much less obvious, and their existence has yet to be fully
understood.

Solitons in Eq. (\ref{e:NLS}) are of the form
\[
\Psi(x, t)=e^{-i\mu t}\psi(x),
\]
where $\mu$ is a real propagation constant, and $\psi(x)$ is a
localized function satisfying the equation
\[
\psi_{xx}+ \mu \hspace{0.04cm} \psi +V(x) \hspace{0.04cm} \psi + \sigma |\psi|^2 \psi =0.
\]
For the complex non-\PT-symmetric potential (\ref{e:V}), these
solitons exist as continuous families, and they can be computed by
various numerical methods such as the squared-operator method and
the Newton-conjugate-gradient method \cite{Yang_book}. To study
their linear stability, we perturb these solitons by infinitesimal
normal modes,
\[
\Psi(x, t)=e^{-i\mu t}\left[\psi(x)+f_1(x) e^{\lambda t}+f_2^*(x) e^{\lambda^*t}\right],
\]
where $|f_1|, |f_2| \ll |\psi|$. Substituting this perturbation into
Eq. (\ref{e:NLS}) and linearizing, we obtain a linear-stability
eigenvalue problem
\begin{equation}\label{e:L}
L \left(\begin{array}{c} f_1\\ f_2
\end{array} \right) = \lambda   \left(\begin{array}{c} f_1\\ f_2
\end{array}\right),
\end{equation}
where
\[
L = \left( \begin{array}{c c} L_{11} &  L_{12} \\
L_{12}^* & L_{11}^*  \end{array} \right),
\]
and
\[  \nonumber
L_{11} = i\left[\partial_{xx} + \mu  +  V(x)  + 2 \sigma |\psi|^2\right],  \quad
L_{12} = i\sigma \psi^2.
\]
This eigenvalue problem can be computed by the Fourier collocation
method (for the full spectrum) or the Newton-conjugate-gradient
method (for individual discrete eigenvalues) \cite{Yang_book}. If
eigenvalues with positive real parts exist, the soliton is linearly
(spectrally) unstable; otherwise it is linearly (spectrally) stable.

Symmetry properties of the linear-stability operator $L$ and its
eigenvalues are important since they strongly influence the
stability results. If the potential $V(x)$ is real [i.e., when Eq.
(\ref{e:NLS}) is Hamiltonian], then $L$ satisfies the following two
symmetry relations,
\[  \label{e:sym1}
L^*=\sigma_1 L \sigma_1^{-1},
\]
\[  \label{e:sym2}
L^\dagger = -\sigma_3 L \sigma_3^{-1},
\]
where the superscript $\dagger$ represents the Hermitian (conjugate
transpose) of a matrix operator, and
\[ \nonumber
\sigma_1=\left[\begin{array}{cc} 0 & 1 \\ 1 & 0\end{array}\right], \quad
\sigma_3=\left[\begin{array}{cc} 1 & 0 \\ 0 & -1\end{array}\right]
\]
are the first and third Pauli spin matrices. The similarity relation
(\ref{e:sym1}) shows that $L^*$ and $L$ share the same spectrum.
Then, since the spectrum of $L^*$ is also the complex conjugate of
the spectrum of $L$, we see that eigenvalues of $L$ must come in
conjugate pairs $(\lambda, \lambda^*)$. The symmetry relation
(\ref{e:sym2}) shows that the spectrum of $L^\dagger$ is negative of
the spectrum of $L$. Since the spectrum of $L^\dagger$ is also
complex conjugate of the spectrum of $L$, eigenvalues of $L$ then
must come in pairs of $(\lambda, -\lambda^*)$. Combining these two
eigenvalue symmetries, we conclude that for real potentials
(Hamiltonian systems), complex eigenvalues of $L$ must come in
quartets $(\lambda, -\lambda, \lambda^*, -\lambda^*)$, which is a
well-known fact. In the special case when the eigenvalue $\lambda$
is real or purely-imaginary, this quartet symmetry reduces to a pair
symmetry $(\lambda, -\lambda)$.

If the potential $V(x)$ is complex but \PT-symmetric, then the
symmetry relation (\ref{e:sym1}) persists, but the other relation
(\ref{e:sym2}) breaks down. In this case, if the soliton $\psi(x)$
is also \PT-symmetric, i.e., $\psi^*(x)=\psi(-x)$, then another
symmetry relation
\[ \label{e:sym3}
L^*=-{\cal P}L{\cal P}^{-1}
\]
is valid, where ${\cal P}$ is the parity operator, i.e., ${\cal P}
f(x) \equiv f(-x)$. Utilizing the two symmetry relations
(\ref{e:sym1}) and (\ref{e:sym3}) and repeating the above arguments,
we conclude that for \PT-symmetric solitons in \PT-symmetric
potentials, complex eigenvalues of $L$ must also come in quartets
$(\lambda, -\lambda, \lambda^*, -\lambda^*)$. This fact has been
pointed out in \cite{Yang_2Dbreaking} although it may not be widely
known.

If the potential $V(x)$ is complex and non-\PT-symmetric, the
symmetry relation (\ref{e:sym1}) still holds, but we cannot see any
additional symmetry for $L$. This suggests that in this case,
complex eigenvalues of $L$ may only appear in conjugate pairs
$(\lambda, \lambda^*)$, but not in quartets. However, a remarkable
discovery from our numerics in the later text is that for
non-\PT-symmetric potentials of the form (\ref{e:V}), eigenvalues of
$L$ still appear as quartets $(\lambda, -\lambda, \lambda^*,
-\lambda^*)$, just as in Hamiltonian and \PT-symmetric systems. This
quartet eigenvalue symmetry has important consequences on the
stability of solitons, which will be described later in this
article.

Our stability analysis will be performed by numerically computing
the spectrum of the linear-stability operator $L$. This spectrum
will be obtained by the Fourier-collocation method \cite{Yang_book}.
Discrete eigenvalues in this spectrum are further checked by the
Newton-conjugate-gradient method \cite{Yang_book}. Both methods can
yield eigenvalues with accuracy of $10^{-10}$ or higher. These
linear-stability results will also be corroborated by direct
evolution simulations of these solitons under initial random-noise
perturbations using the pseudo-spectral method \cite{Yang_book}. In
our numerical examples, we take
\[  \label{e:g}
g(x)=\tanh2(x+2.5)-\tanh(x-2.5),
\]
which is an asymmetric single-hump function. For this choice of
$g(x)$, a phase transition occurs at $\gamma=\gamma_c \approx
-0.1806$, where the linear spectrum of the potential is all-real
when $\gamma>\gamma_c$ and becomes partially complex when
$\gamma<\gamma_c$.

\section{Linear stability of solitons below phase transition}

\begin{figure}[tbh!]
\includegraphics[width=0.48\textwidth]{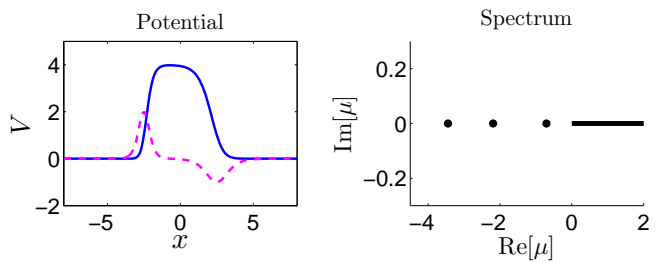}

\smallskip
\caption{(Color online) The complex potential (\ref{e:V}) with
$g(x)$ given in (\ref{e:g}) and $\gamma=0$ (left) and its linear
spectrum (right).} \label{f:fig1}
\end{figure}

First, we consider the linear stability of soliton families below
phase transition. For this purpose, we take $\gamma=0$. The
resulting potential $V(x)$ and its linear spectrum are displayed in
Fig.~\ref{f:fig1}. Notice that the real part of $V(x)$ is not even,
and its imaginary part not odd, thus this potential is
non-\PT-symmetric, but its spectrum is all-real. In addition, this
spectrum contains three discrete eigenvalues,
\[ \nonumber
\mu_1\approx -3.4484, \quad \mu_2\approx -2.1899, \quad \mu_3 \approx -0.7044.
\]

\begin{figure}[tbh!]
\hspace{0.38cm}\includegraphics[width=0.458\textwidth]{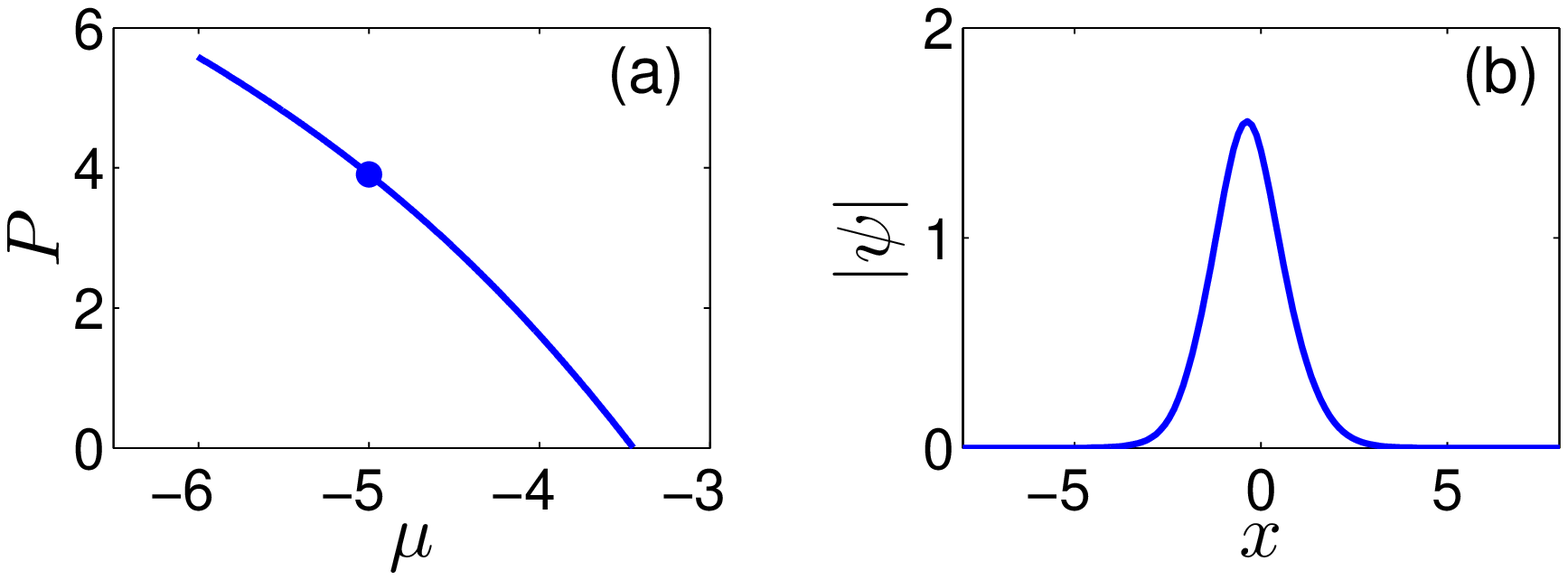}

\includegraphics[width=0.48\textwidth]{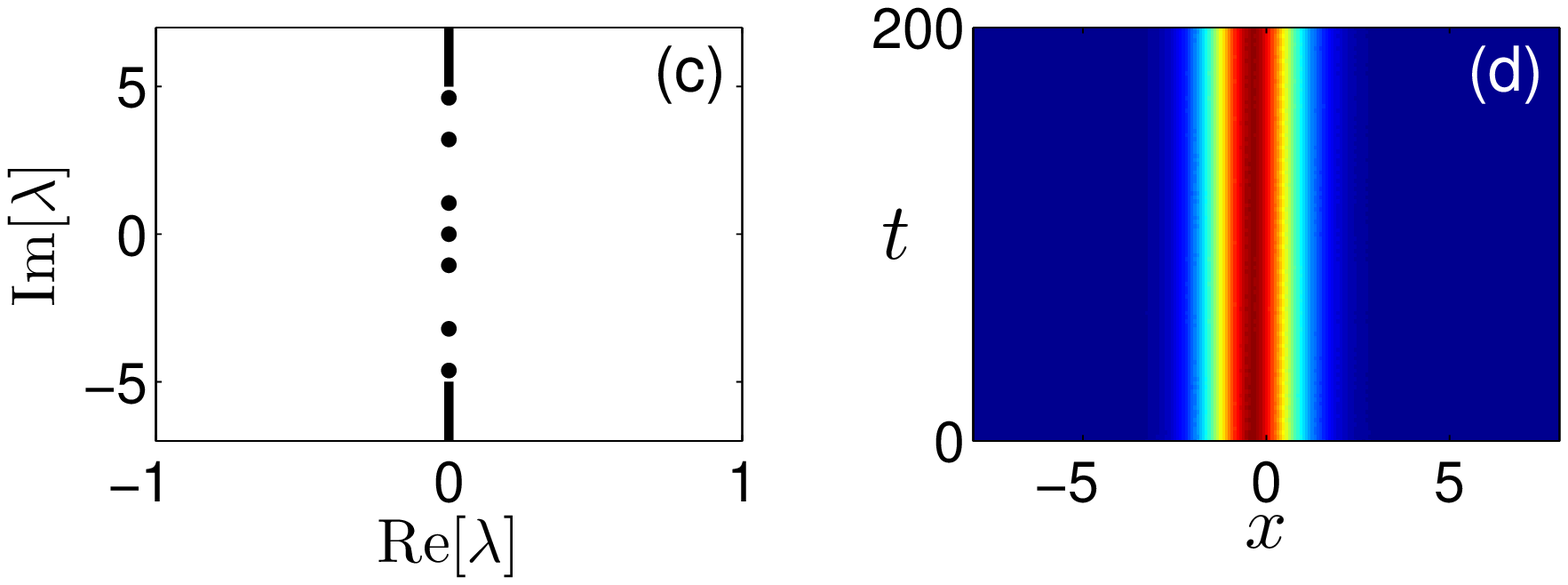}
\smallskip
\caption{(Color online) (a) Power curve of fundamental solitons
(bifurcating from the eigenvalue $\mu_1$) below phase transition
under focusing nonlinearity. (b) Amplitude profile of the soliton at
the marked point of the power curve. (c) Linear-stability spectrum
for the soliton in (b). (d) Time evolution of the soliton in (b)
under 1\% random-noise perturbations. } \label{f:fig2}
\end{figure}

Continuous families of solitons can bifurcate out from each of these
three discrete eigenvalues under either sign of nonlinearity. We
first consider the soliton family bifurcating from the first
(lowest) eigenvalue $\mu=\mu_1$ under focusing nonlinearity
($\sigma=1$). Defining the power of a soliton as
$P=\int_{-\infty}^\infty |\psi|^2dx$, the power curve of this
soliton family is plotted in Fig.~\ref{f:fig2}(a). At the marked
point of this power curve (with $\mu=-5$), the amplitude profile of
the soliton is displayed in Fig.~\ref{f:fig2}(b). It is seen that
this amplitude profile is single-humped. Since this soliton family
bifurcates from the lowest eigenvalue $\mu_1$, we call this family
of solitons \emph{fundamental solitons}. We have computed the
linear-stability spectra for these fundamental solitons, and found
that their eigenvalues all lie on the imaginary axis. Thus, these
fundamental solitons are linearly stable. As an example, the
linear-stability spectrum for the soliton of Fig.~\ref{f:fig2}(b) is
shown in Fig.~\ref{f:fig2}(c). This spectrum consists of three pairs
of discrete non-zero eigenvalues and the continuous spectrum, all on
the imaginary axis. Time evolution of this soliton for 200 time
units under initial 1\% random noise perturbations is plotted in
Fig.~\ref{f:fig2}(d). It is seen that this soliton is robust against
perturbations, consistent with its linear-stability result.

\begin{figure}[tbh!]
\hspace{0.38cm}\includegraphics[width=0.455\textwidth]{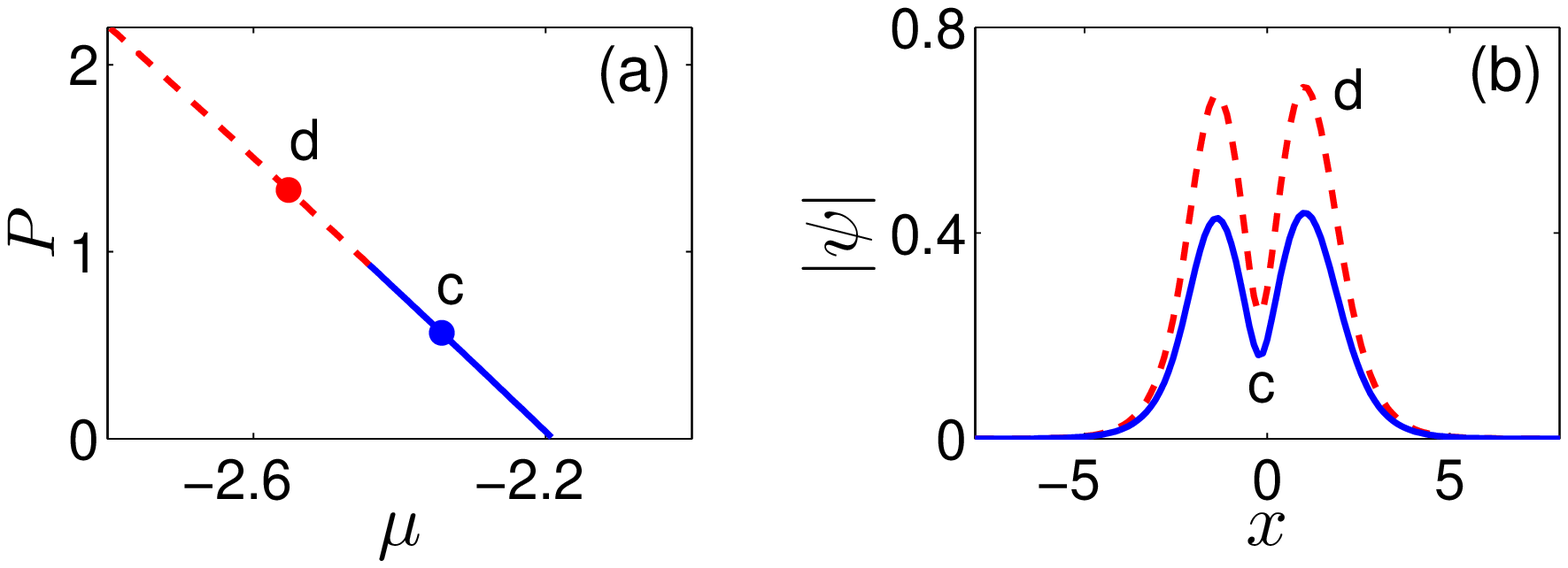}

\vspace{0.1cm}
\includegraphics[width=0.48\textwidth]{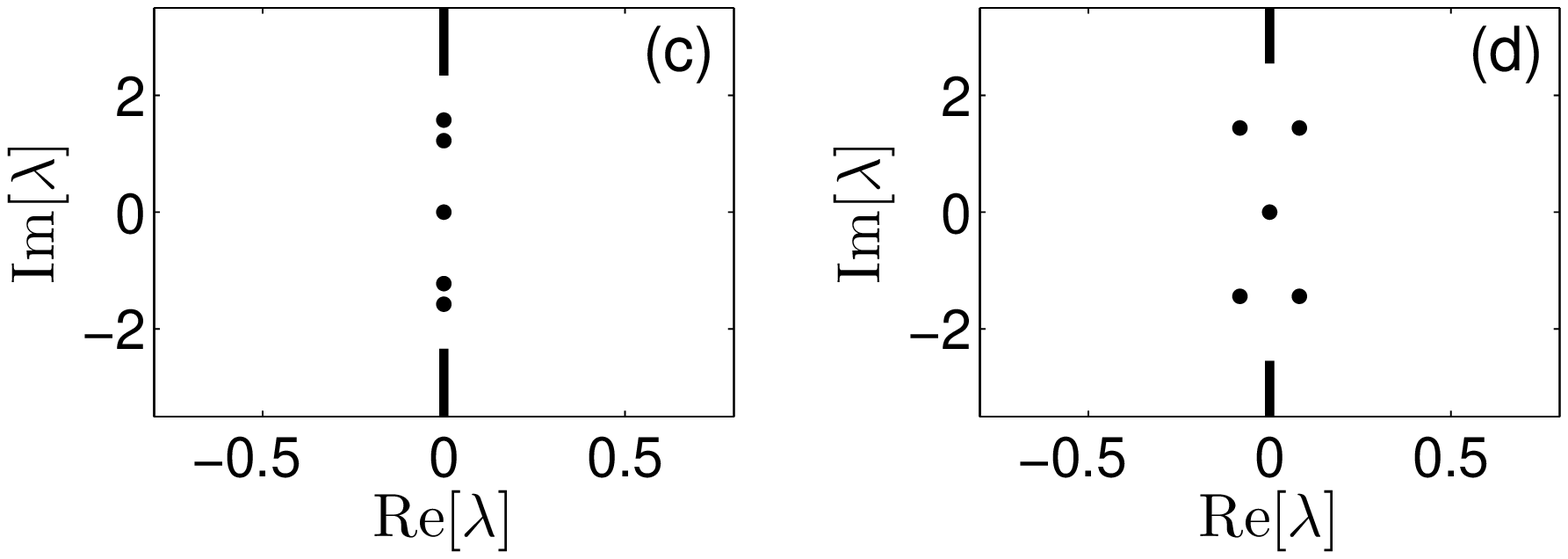}

\hspace{0.2cm}
\includegraphics[width=0.465\textwidth]{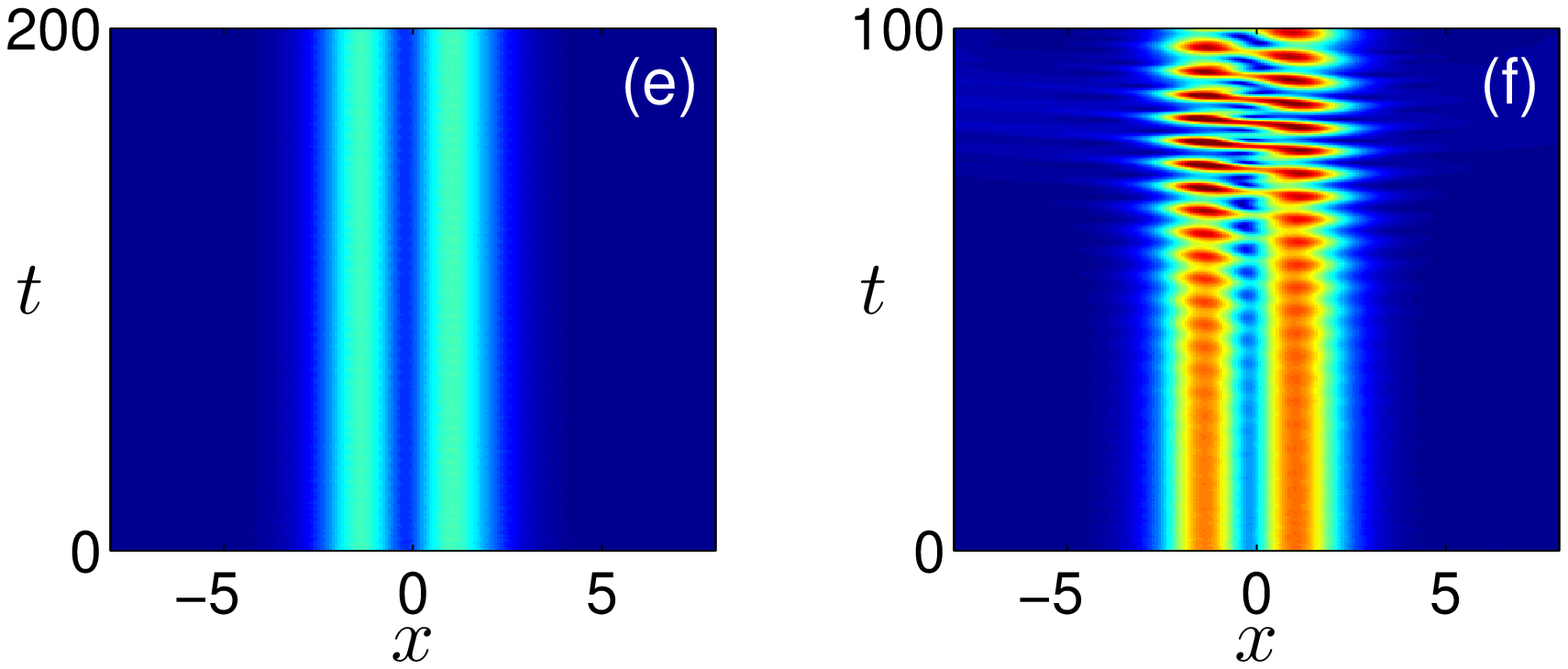}
\smallskip
\caption{(Color online) (a) Power curve of excited-state solitons
(bifurcating from the eigenvalue $\mu_2$) below phase transition
under focusing nonlinearity (solid blue indicates stability, and
dashed red indicates instability). (b) Amplitude profiles of
solitons at the marked points `c, d' of the power curve (lower for
`c' and upper for `d'). (c, d) Linear-stability spectra for the
lower and upper solitons in panel (b) respectively. (e, f) Time
evolutions of the lower and upper solitons in panel (b) under 1\%
random-noise perturbations respectively.  } \label{f:fig3}
\end{figure}

Next, we consider the soliton family bifurcating from the second
eigenvalue $\mu=\mu_2$ under focusing nonlinearity. The power curve
of this soliton family is displayed in Fig.~\ref{f:fig3}(a). At the
marked points `c, d' of this power curve (with $\mu=-2.34$ and
$-2.55$ respectively), amplitude profiles of the solitons are shown
in Fig.~\ref{f:fig3}(b). These profiles are double-humped,
indicating that this family of solitons are excited states. At low
powers, these solitons are linearly stable. This is evidenced by the
linear-stability spectrum shown in Fig.~\ref{f:fig3}(c) for the
lower-power soliton in Fig.~\ref{f:fig3}(b), where all eigenvalues
are purely imaginary. However, at higher powers, these excited-state
solitons become linearly unstable. This can be seen from the
linear-stability spectrum in Fig.~\ref{f:fig3}(d) for the
higher-power soliton in Fig.~\ref{f:fig3}(b). In this spectrum, a
quartet of complex eigenvalues appear, creating oscillatory
instability. To corroborate these linear-stability results, we have
simulated the evolutions of the two solitons in Fig.~\ref{f:fig3}(b)
under 1\% random-noise perturbations, and the simulation results are
displayed in Fig.~\ref{f:fig3}(e,f). The panel (e) shows robust
(stable) propagation, while the panel (f) shows the onset of
oscillatory instability, consistent with the linear-stability
results.

The change of linear stability in this family of excited-state
solitons occurs at $\mu=\mu_c\approx -2.440$, where $P_c\approx
0.925$. This instability arises when two pairs of imaginary
eigenvalues bifurcate into a complex quartet. Below the critical
power $P_c$, the stability spectrum of the soliton contains two
pairs of imaginary eigenvalues $(\pm i\omega_1, \pm i\omega_2)$ [see
Fig.~\ref{f:fig3}(c)]. As the soliton's power increases to $P_c$,
eigenvalues $\pm i\omega_1$ and $\pm i\omega_2$ approach each other
on the imaginary axis. At the critical power $P_c$, these imaginary
eigenvalues coalesce and form exceptional points with geometric
multiplicity one and algebraic multiplicity two. Above the critical
power $P_c$, these exceptional points bifurcate off the imaginary
axis, creating a quartet of complex eigenvalues (and hence
oscillatory instability) [see Fig.~\ref{f:fig3}(d)]. This
instability mechanism is remarkably similar to Hamiltonian-Hopf
bifurcations in Hamiltonian systems [such as when the potential
$V(x)$ in Eq. (\ref{e:NLS}) is real] \cite{Goodman,Yang_HH}, even
though the present system is non-Hamiltonian. Thus, we call this
change of linear stability \emph{pseudo-Hamiltonian-Hopf
bifurcation}.

We note that pseudo-Hamiltonian-Hopf bifurcations can appear in
\PT-symmetric systems as well. This fact has not been reported in
the literature yet, but we have seen it in our numerics of
\PT-symmetric systems [such as in Eq. (\ref{e:NLS}) with a
\PT-symmetric potential].

A remarkable feature in the above stability results is that,
although Eq. (\ref{e:NLS}) in our consideration is non-Hamiltonian
and non-\PT-symmetric, and the gain and loss in the system are
rather large, stability behaviors of solitons in our system are
analogous to those in Hamiltonian systems. Examples include the
stability of fundamental solitons and pseudo-Hamiltonian-Hopf
bifurcations of excited-state solitons
\cite{Kapitula_book,Goodman,Yang_HH}. However, differences in
behaviors between our system and Hamiltonian systems also exist. For
instance, in our system, low-amplitude fundamental solitons can be
linearly unstable, and pseudo-Hamiltonian-Hopf bifurcations can
occur on fundamental solitons at higher powers. These phenomena
would not happen in Hamiltonian systems. These differences will be
shown in the next section.

\section{Linear stability of solitons above phase transition}

\begin{figure}[tbh!]
\includegraphics[width=0.48\textwidth]{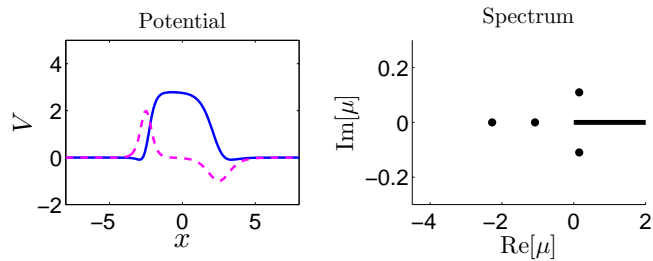}

\smallskip
\caption{(Color online) The complex potential (\ref{e:V}) with
$g(x)$ given in (\ref{e:g}) and $\gamma=-0.3$ (left) and its linear
spectrum (right). }  \label{f:fig4}
\end{figure}

Now we consider the stability of solitons above phase transition.
For this purpose, we take $\gamma=-0.3$. The resulting potential
(\ref{e:V}) is displayed in Fig.~\ref{f:fig4}. Clearly, this
potential is also non-\PT-symmetric. When compared to the potential
in Fig.~\ref{f:fig1}, the real part of this potential is
significantly lower, while its imaginary part remains the same. The
linear spectrum of this potential, displayed also in
Fig.~\ref{f:fig4}, shows the presence of a pair of complex
eigenvalues, indicating that this potential is above phase
transition. It is noted that for complex potentials of the form
(\ref{e:V}), complex eigenvalues in the linear spectrum of the
potential appear as conjugate pairs $(\mu, \mu^*)$ \cite{PRA2016}.
This eigenvalue symmetry is clearly visible in Fig.~\ref{f:fig4}.

\begin{figure}[tbh!]
\hspace{0.38cm}\includegraphics[width=0.455\textwidth]{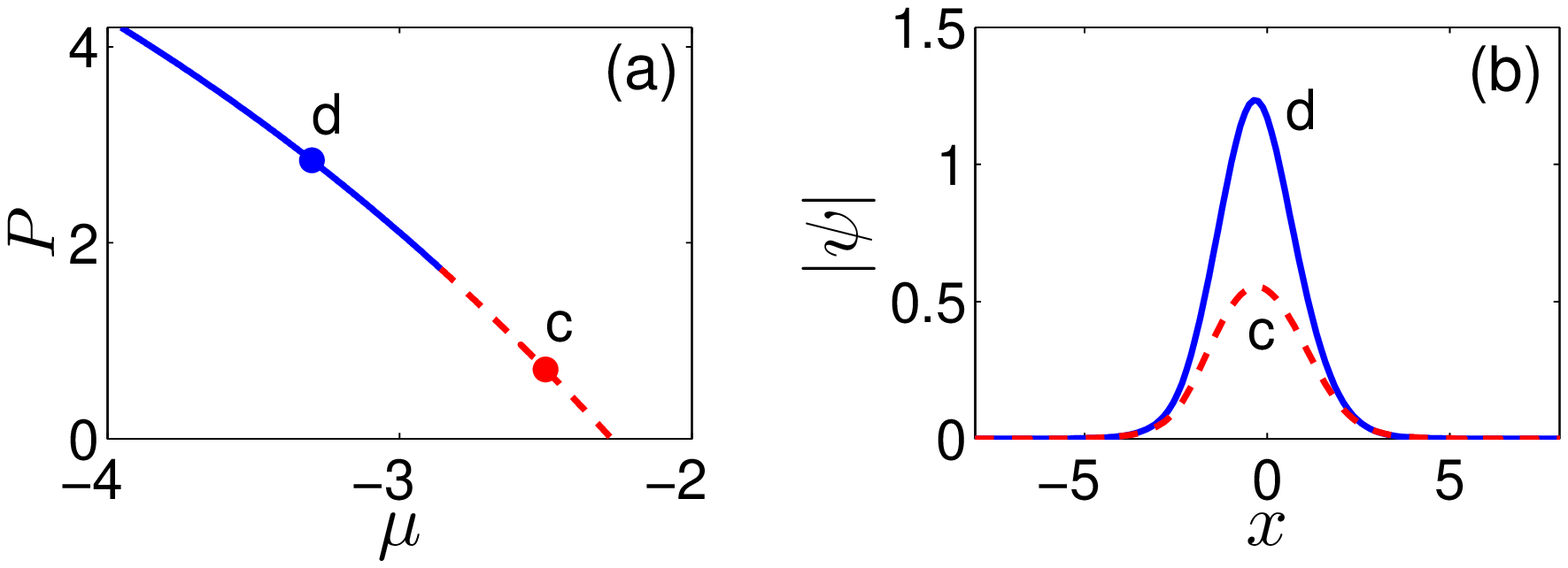}

\vspace{0.1cm}
\includegraphics[width=0.48\textwidth]{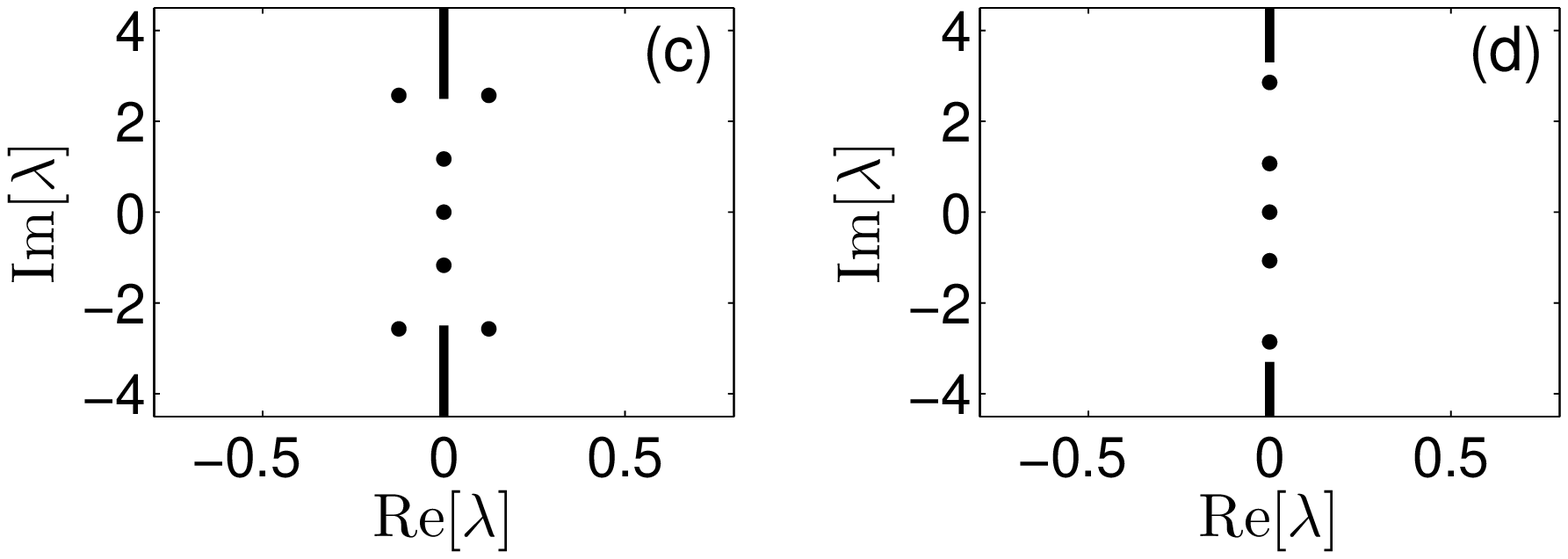}

\includegraphics[width=0.48\textwidth]{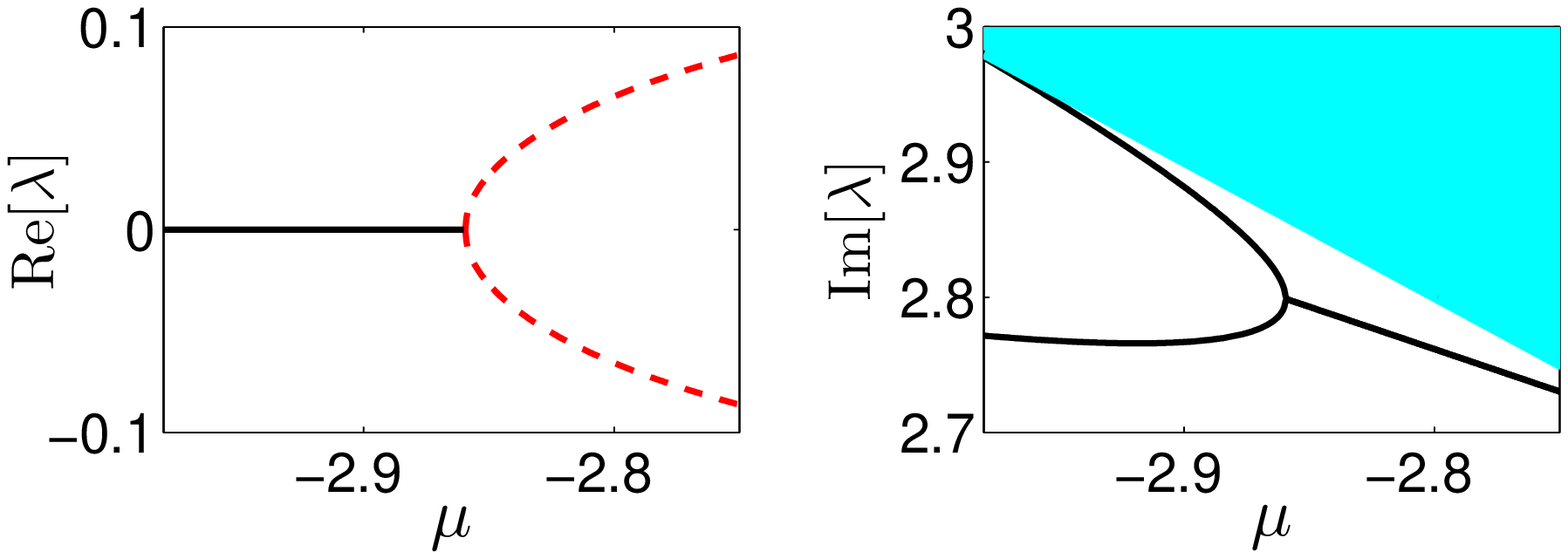}
\smallskip
\caption{(Color online) (a) Power curve of fundamental solitons
(bifurcating from the eigenvalue $\mu_1$) above phase transition
under focusing nonlinearity (solid blue indicates stability, and
dashed red indicates instability). (b) Amplitude profiles of
solitons at the marked points `c, d' of the power curve (lower for
`c' and upper for `d'). (c, d) Linear-stability spectra for the
lower and upper solitons in panel (b) respectively. (e, f) Real and
imaginary parts of linear-stability eigenvalues $\lambda$ versus the
propagation constant $\mu$ (the continuous spectrum is shown in
light blue). } \label{f:fig5}
\end{figure}

In addition to the pair of complex eigenvalues, the spectrum in
Fig.~\ref{f:fig4} also contains two discrete real eigenvalues,
\[  \nonumber
\mu_1\approx -2.2740, \quad \mu_2\approx -1.0787.
\]
From these real eigenvalues, soliton families can bifurcate out
under either sign of nonlinearity. Here, we consider the soliton
family bifurcating from the lowest eigenvalue $\mu_1$ under focusing
nonlinearity ($\sigma=1$). The power curve of this soliton family is
plotted in Fig.~\ref{f:fig5}(a), and the solitons at the marked
points `c, d' of the power curve, with $\mu=-2.5$ and $-3.3$
respectively, are shown in Fig.~\ref{f:fig5}(b). These solitons have
a single-hump amplitude profile, and are fundamental solitons due to
their bifurcation from the lowest eigenvalue $\mu_1$.

At low powers, these solitons are linearly unstable because the
underlying potential is above phase transition. This is evidenced in
Fig.~\ref{f:fig5}(c), where the linear-stability spectrum for the
lower-power soliton in Fig.~\ref{f:fig5}(b) is displayed. This
instability is due to a quartet of complex eigenvalues $(\lambda_0,
\lambda_0^*, -\lambda_0, -\lambda_0^*)$, which are directly related
to the pair of complex eigenvalues in the linear spectrum of the
potential in Fig.~\ref{f:fig4}. However, at higher powers, these
solitons become linearly stable. This is evidenced in
Fig.~\ref{f:fig5}(d), where the linear-stability spectrum for the
higher-power soliton of Fig.~\ref{f:fig5}(b) is plotted. All
eigenvalues in this spectrum are on the imaginary axis, indicating
this soliton is linearly stable. The change of stability occurs at
$\mu=\mu_c\approx -2.86$, where $P_c\approx 1.74$.

The reason for this stabilization of solitons at higher powers is
that, as the power increases toward the critical power $P_c$, the
quartet of complex eigenvalues  $(\lambda_0, \lambda_0^*,
-\lambda_0, -\lambda_0^*)$ move toward the imaginary axis. At the
critical power $P_c$, these complex eigenvalues collide on the
imaginary axis and create a pair of exceptional points. Above the
critical power, these exceptional points split along the imaginary
axis and become two pairs of imaginary eigenvalues $\{\pm i\omega_1,
\pm i \omega_2\}$, thus the solitons become linearly stable. This
stabilization process is more clearly depicted in
Fig.~\ref{f:fig5}(e,f), where the real and imaginary parts of the
relevant linear-stability eigenvalues are plotted versus the
propagation constant $\mu$. This stabilization is a reverse
pseudo-Hamiltonian-Hopf bifurcation as the power rises.

In this example, fundamental solitons at low powers are linearly
unstable, and a reverse pseudo-Hamiltonian-Hopf bifurcation is seen.
These phenomena will not occur in Hamiltonian systems, such as Eq.
(\ref{e:NLS}) with a real potential. In such Hamiltonian systems,
fundamental solitons at low amplitudes are always linearly and
nonlinearly stable because their Hamiltonian-Krein index is zero
\cite{Kapitula_book}. In addition, Hamiltonian-Hopf bifurcations
cannot occur on fundamental solitons of any powers, because there
are no imaginary eigenvalues with negative Krein signatures, but
such imaginary eigenvalues are necessary for Hamiltonian-Hopf
bifurcations \cite{Kapitula_book}.

\section{Quartet eigenvalue symmetry and its consequences}

The most surprising finding of the above stability analysis is that,
linear-stability eigenvalues of solitons in non-\PT-symmetric
potentials (\ref{e:V}) appear in quartets $(\lambda, \lambda^*,
-\lambda, -\lambda^*)$, i.e., if $\lambda$ is an eigenvalue of the
operator $L$, so are $\lambda^*, -\lambda$ and $-\lambda^*$. As we
have pointed out in Sec. \ref{s:sec2}, in non-\PT-symmetric
potentials (\ref{e:V}), linear-stability eigenvalues still appear in
pairs $(\lambda, \lambda^*)$, i.e., the spectrum is symmetric with
respect to the real axis. But we cannot see another symmetry of the
operator $L$ which assures the appearance of eigenvalues in
$(\lambda, -\lambda^*)$ pairs, i.e., the spectrum's symmetry with
respect to the imaginary axis. Because of this, we do not anticipate
the quartet eigenvalue symmetry in the spectrum of $L$. However, our
numerical results in the earlier text show that these eigenvalues do
come in quartets of $(\lambda, \lambda^*, -\lambda, -\lambda^*)$,
which is very remarkable.

\begin{figure}[tbh!]
\includegraphics[width=0.48\textwidth]{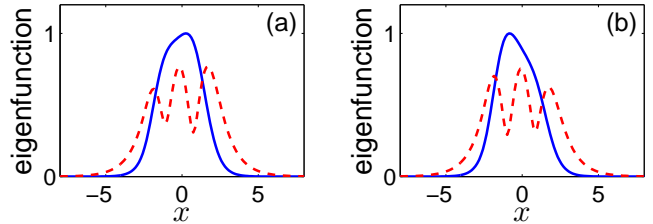}

\smallskip
\caption{(Color online) Eigenfunctions $(f_1, f_2)$ for the upper
two complex eigenvalues in the spectrum of Fig.~\ref{f:fig3}(d),
with $\lambda=\lambda_1$ in (a) and $\lambda=\lambda_2$ in (b),
where $\lambda_{1, 2}$ are given in (\ref{e:lam1})-(\ref{e:lam2}).
Solid blue lines are $|f_1|$ and dashed red lines $|f_2|$.}
\label{f:fig6}
\end{figure}

The visual evidence of this quartet eigenvalue symmetry can already
be seen in the stability spectra of Figs. 2, 3 and 5, where the
spectra are always symmetric with respect to both the real and
imaginary axes. Here, we establish this eigenvalue symmetry
quantitatively. Since the $(\lambda, \lambda^*)$ symmetry is already
known, we focus on the $(\lambda, -\lambda^*)$ symmetry below. To
establish this latter symmetry, we first consider the two upper
complex eigenvalues in Fig.~\ref{f:fig3}(d). Numerical computations
give these two eigenvalues (accurate to all twelve digits) as
\[ \label{e:lam1}
\lambda_1=-0.08220738069... + 1.43969109965...i,
\]
\[ \label{e:lam2}
\lambda_2=\hspace{0.2cm} 0.08220738069... + 1.43969109965...i.
\]
Clearly, $\lambda_2=-\lambda_1^*$ to numerical accuracy. Using
multi-precision computation, we have further checked that
$\lambda_2$ and $-\lambda_1^*$ match each other to many more digits.
The eigenfunctions $(f_1, f_2)$ of these two eigenvalues are plotted
in Fig. \ref{f:fig6}. Notice that these two eigenfunctions are not
related to each other by any obvious symmetry. However, their
eigenvalues are related as $\lambda_2=-\lambda_1^*$, which is quite
intriguing.

As another example, we consider the two upper complex eigenvalues in
the spectrum of Fig.~\ref{f:fig5}(c). Numerical computations give
these two eigenvalues (accurate to all twelve digits) as
\[ \nonumber
\lambda_1= -0.12447936624... + 2.57218047717...i,
\]
\[ \nonumber
\lambda_2=\hspace{0.2cm} 0.12447936624... + 2.57218047717...i.
\]
Again, $\lambda_2=-\lambda_1^*$ to numerical accuracy.

We have examined the other eigenvalues in the spectra of Figs. 2, 3
and 5, and found them to lie exactly on the imaginary axis (to high
numerical accuracy). Thus, these spectra are indeed symmetric with
respect to both the real and imaginary axes, confirming the quartet
eigenvalue symmetry of $(\lambda, \lambda^*, -\lambda, -\lambda^*)$.

In addition to the stability spectra in Figs. 2, 3 and 5, we have
examined the spectra of solitons in other non-\PT-symmetric
potentials of the form (\ref{e:V}), which are not included in this
article. Those stability spectra respect the quartet eigenvalue
symmetry as well.

This quartet eigenvalue symmetry in the linear-stability spectrum
has far reaching consequences on the linear-stability behaviors of
solitons. First, it assures the linear stability of low-power
fundamental solitons (bifurcating from the lowest discrete
eigenvalue of the potential) as long as the potential is below phase
transition. This follows from the fact that a potential which is
below phase transition has all-real spectra and its discrete
non-zero eigenvalues can be assumed to be all simple (which is the
generic case). Then, in the limit of zero amplitude of fundamental
solitons, the linear-stability spectrum (of operator $L$) is purely
imaginary, and all discrete non-zero eigenvalues of $L$ are simple.
In addition, no discrete eigenvalues are embedded inside the
continuous spectrum. When the amplitude of the soliton is non-zero
but small, by virtue of the eigenvalue continuity and quartet
eigenvalue symmetry, the simple discrete imaginary eigenvalues of
$L$ cannot move off the imaginary axis. Meanwhile, the zero
eigenvalue and the continuous spectrum do not change. Thus, the
spectrum remains on the imaginary axis, and low-amplitude
fundamental solitons are linearly stable below phase transition.
This analytically explains our numerical findings for low-power
solitons in Fig.~\ref{f:fig2}.

By a similar argument, we can also show that, in the presence of
this quartet eigenvalue symmetry, low-power excited-state solitons
(bifurcating from the higher discrete real eigenvalues $\mu_k$ of
the potential with $k>1$) are also linearly stable if the potential
is below phase transition, and none of $i(\mu_k-\mu_j)$ ($j\ne k$)
is embedded inside the continuous spectrum of operator $L$ when
$\psi(x)=0$. For the example in Fig.~\ref{f:fig3}, the latter
condition means $|\mu_2-\mu_j|<|\mu_2|$ ($j=1, 3$), which is
satisfied. This analytically explains the linear stability of
low-power excited-state solitons below phase transition in
Fig.~\ref{f:fig3}.

Another consequence of this quartet eigenvalue symmetry is that it
makes pseudo-Hamiltonian-Hopf bifurcation possible in the
non-Hamiltonian system (\ref{e:NLS}). If the linear-stability
spectrum contains two pairs of simple imaginary eigenvalues $\{\pm
i\omega_1, i\omega_2\}$ for a certain soliton, then when the
propagation constant of the soliton continuously changes, these
simple imaginary eigenvalues have to stay on the imaginary axis due
to the quartet eigenvalue symmetry. In this case, if these
eigenvalues move toward each other and collide ($\omega_1\to
\omega_2$), they could leave the imaginary axis and become a quartet
of complex eigenvalues, creating a pseudo-Hamiltonian-Hopf
bifurcation. This is exactly what happens in Figs.~\ref{f:fig3} and
Fig.~\ref{f:fig5}.

One more consequence of this quartet eigenvalue symmetry is that it
closely mimics that of solitons in Hamiltonian systems and of
\PT-symmetric solitons in \PT-symmetric systems (see Sec.
\ref{s:sec2}). This implies that non-\PT-symmetric solitons in
complex potentials (\ref{e:V}) are likely to share many stability
properties of those other systems, as the results of this paper have
shown.

\section{Summary and discussion}
In this paper, we have numerically analyzed the linear stability of
soliton families in 1D NLS equations (\ref{e:NLS}) with
non-\PT-symmetric complex potentials (\ref{e:V}). We have shown that
these solitons can be linearly stable in a wide range of parameter
values both below and above phase transition. More importantly, we
have discovered that linear-stability eigenvalues of these solitons
appear in quartets $(\lambda, -\lambda, \lambda^*, -\lambda^*)$,
similar to conservative systems and \PT-symmetric systems. This
quartet eigenvalue symmetry is very surprising for non-\PT-symmetric
systems, and it facilitates the existence of stable solitons and
makes their pseudo-Hamiltonian-Hopf bifurcation possible.

A question closely related to the subject of this paper is the
linear stability of asymmetric solitons in \PT-symmetric potentials.
Earlier work has shown that in \PT-symmetric potentials of the same
form (\ref{e:V}) [where $g(x)$ is now an even function], symmetry
breaking of solitons can occur \cite{Yang_OL14}. That is, from the
base branch of \PT-symmetric solitons, a branch of asymmetric
solitons can bifurcate out. The linearization operator $L$ of these
asymmetric solitons only admits the symmetry (\ref{e:sym1}), as far
as one can see, similar to the present case. This $L$ symmetry only
assures its eigenvalue symmetry of $(\lambda, \lambda^*)$. However,
our numerical studies (not shown in this article) have revealed
that, their linear-stability eigenvalues also appear in quartets of
$(\lambda, -\lambda, \lambda^*, -\lambda^*)$, closely resembling the
findings in this article for non-\PT-symmetric potentials
(\ref{e:V}). This quartet eigenvalue symmetry for asymmetric
solitons in \PT-symmetric potentials (\ref{e:V}) is equally
surprising.

Another question closely related to the subject of this paper is the
linear stability of two-dimensional solitons in non-\PT-symmetric
complex potentials. Earlier work has shown that in two-dimensional
non-\PT-symmetric complex potentials of certain forms, continuous
families of solitons can also bifurcate out from linear modes
\cite{Yang_2Dbreaking}. Our numerical studies (not included in this
article) have found that linear-stability eigenvalues of those 2D
solitons only possess the conjugate-pair symmetry of $(\lambda,
\lambda^*)$, but NOT the quartet symmetry of $(\lambda, -\lambda,
\lambda^*, -\lambda^*)$. This result echoes that for
linear-stability eigenvalues of asymmetric solitons in 2D
partially-\PT-symmetric potentials \cite{Yang_2Dbreaking}. These
results indicate that the quartet eigenvalue symmetry depends on the
spatial dimension of the problem.

Many questions are still open regarding the findings of this paper.
The most important question is why the quartet eigenvalue symmetry
appears for linear-stability eigenvalues of solitons in
non-\PT-symmetric complex potentials (\ref{e:V}). A related question
is why this quartet eigenvalue symmetry also appears for asymmetric
solitons in \PT-symmetric potentials (\ref{e:V}). In both cases, the
linear-stability operator $L$ seems to only admit the symmetry
(\ref{e:sym1}). Whether this operator also admits another hidden
symmetry which assures the eigenvalue symmetry of $(\lambda,
-\lambda^*)$ is an open question. If such a hidden $L$ symmetry
cannot be found, then how to explain the quartet eigenvalue symmetry
of $L$ remains to be seen.

Another important open question concerns the nonlinear stability of
solitons in non-\PT-symmetric complex potentials (\ref{e:V}). In
this paper, our focus was the linear (spectral) stability of these
solitons, and we showed that these solitons can be linearly stable
in a wide range of parameter values. However, it is well known that
solitons can be nonlinearly unstable even if they are linearly
stable. For instance, in Hamiltonian systems, linearly-stable
solitons are nonlinearly unstable if their linear-stability spectrum
contains imaginary eigenvalues with negative Krein signatures
\cite{Kapitula_book,Peli_oscillation_2015,Yang_HH} (these modes with
negative Krein signatures are often called negative-energy modes in
the physics literature \cite{Morrison}). In our non-Hamiltonian
system (\ref{e:NLS}), whether these linearly-stable solitons are
nonlinearly stable or not is still an open question. One might think
that robust evolution simulations of such solitons under
random-noise perturbations in Figs. \ref{f:fig2}(d) and
\ref{f:fig3}(e) should indicate that those solitons are also
nonlinearly stable. Such a conclusion is too hasty, since nonlinear
instability is often slower and may take longer time to develop
\cite{Peli_oscillation_2015,Yang_HH}. Regarding the issue of
nonlinear stability, we should add that this question is also open
for solitons in \PT-symmetric systems. A little progress has been
made in this direction. In a \emph{linear} Schr\"odinger equation
with a \PT-symmetric potential, which arises when considering the
stability of the zero state in Eq. (\ref{e:NLS}), a \PT-Krein
signature theory was developed recently \cite{PhysicaD2016}. This
theory can be readily extended to the linear Schr\"odinger equation
with a non-\PT-symmetric potential of the form (\ref{e:V})
\cite{PhysicaD2016,PRA2016}. Whether a similar theory can be
developed for solitons in the non-Hamiltonian system (\ref{e:NLS})
remains to be seen.

\vspace{0.2cm} \noindent \textbf{Acknowledgments:} This work was
supported in part by the Air Force Office of Scientific Research
(Grant USAF 9550-12-1-0244) and the National Science Foundation
(Grant DMS-1311730).


\end{document}